\documentclass[12pt]{article}

\usepackage{amsmath,amsfonts,amssymb,alltt}
\usepackage{epsfig}
\usepackage{graphics}
\usepackage{graphicx}

\usepackage{amscd}
\usepackage{latexsym}
\usepackage{graphicx}
\usepackage{xcolor,amsmath,amssymb,amsbsy,tikz,graphicx}

\usepackage{cite}

\usepackage{amssymb}
\usepackage{amsmath}
\usepackage{amscd}
\usepackage{latexsym}
\usepackage{graphicx}

\newcommand{\be}{\begin{equation}}
\newcommand{\ee}{\end{equation}}
\newcommand{\Dlt}{\Delta}

\newcommand{\ra}{\rightarrow}

\newcommand{\bt}{\beta}

\newcommand{\gm}{\gamma}

\newcommand{\Gm}{\Gamma}

\newcommand{\prt}{\partial}

\begin{document}

\begin{center}
{\Large {\bf Resolving the Problem of Multiple Control Parameters \\ [2mm]
in Optimized Borel-Type Summation }} \\ [5mm]

 V.I. Yukalov$^{1,2}$ and S. Gluzman$^3$  \\ [3mm]

{\it $^1$Bogolubov Laboratory of Theoretical Physics, \\
Joint Institute for Nuclear Research, Dubna 141980, Russia \\ [2mm]

$^2$Instituto de Fisica de S\~ao Carlos, Universidade de S\~ao Paulo, \\
CP 369, S\~ao Carlos 13560-970, S\~ao Paulo, Brazil  \\ [2mm]

$^3$Materialica + Research Group, Bathurst St. 3000, Apt. 606, \\
Toronto, ON M6B 3B4, Canada } \\ [3mm]

Correspondence: yukalov@theor.jinr.ru 

\end{center}

\vskip 3cm

\begin{abstract}

One of the most often used methods of summing divergent series in physics 
is the Borel-type summation with control parameters improving convergence, 
which are defined by some optimization conditions. The well known annoying
problem in this procedure is the occurrence of multiple solutions for control 
parameters. We suggest a method for resolving this problem, based on the 
minimization of cost functional. Control parameters can be introduced by 
employing the Borel-Leroy or Mittag-Leffler transforms. Also, two novel 
transformations are proposed using fractional integrals and fractional derivatives. 
New cost functionals are advanced, based on lasso and ridge selection criteria, 
and their performance is studied for a number of models. The developed method 
is shown to provide good accuracy for the calculated quantities.  
      
\end{abstract}

\vskip 1cm

{\parindent=0pt


\vskip 1cm

 }  

\section{Introduction}

Solutions to nontrivial problems of physics and applied mathematics are often presented 
in the form of asymptotic expansions in powers of some parameters. These expansions are
usually divergent. In order to induce convergence, the expansions are subject to 
transformations containing control functions or control parameters. For instance, 
control parameters can be introduced through Borel-type transforms. Control functions 
or parameters are defined by optimization conditions governing the convergence of
the transformed expansions. In many cases, the optimization conditions lead to multiple
solutions. Then we confront the problem of choosing among multiple control parameters.

The paper addresses this problem of dealing with multiple control parameters introduced
by Borel-type transformations. For this purpose the optimization conditions are formulated
as the conditions of minimization of cost functionals. Several cost functionals are 
considered and their efficiency is compared for a number of physical examples. 

We pay the major attention to the most difficult problem of finding the large-variable
behavior of a physical quantity represented by a finite and short asymptotic expansion
at small variables. To be precise, we concentrate on the case of physical quantities 
with power-law behavior of variables tending to infinity. We show that even a few 
terms of a series at small variables are sufficient for defining the large-variable 
behavior of the sought physical quantity. The considered cost functionals lead to rather
close results well representing the infinite-variable behavior of the sought function.       

In the power-law problems, there are two quantities to be found, the power and the 
amplitude. Often, the power can be obtained from the scaling arguments, large-variable 
properties of the studied function, renormalization-group techniques, and other methods 
for defining critical indices. If the power is found, then one needs to find the 
amplitude, which is also not a simple task. For convenience, the problem can be 
separated into two parts, definition of the power and of the amplitude. Here, we assume 
that the power can be obtained employing the methods mentioned above and what is left 
is to get an accurate approximation for the amplitude. The main goal of the present 
paper is to elaborate methods allowing for an accurate estimation of the amplitudes 
of power-law functions at large-variables.

\section{Control Functions and Parameters}

Suppose we are looking for a physical quantity $f(x)$ that is a real function of a real 
variable. Usually, the solution to a complicated physical problem is given, by applying 
a kind of perturbation theory, in the form of a truncated asymptotic expansion
\be
\label{1}
f_k(x) = \sum_{n=0}^k a_n x^n \; ,
\ee
valid at asymptotically small variable $x \ra 0$. Being based on this knowledge, we 
need to find out the behavior of the sought physical quantity at asymptotically large 
variable $x \ra \infty$, where it is assumed to enjoy the power-law behavior
\be
\label{2}
f(x) \simeq B x^\bt \qquad (x \ra \infty) \;  .
\ee
For concreteness, let us consider the case, where, from some physical arguments, the 
power $\beta$ is known and we have to find the value of the critical amplitude $B$.

Asymptotic series, corresponding to interesting physical problems, are usually 
divergent \cite{Hardy_1,Kardar_2}. In order to extract information from divergent 
series it is necessary to rearrange them into convergent series by introducing control 
functions or control parameters whose role is to govern the series convergence. 
Symbolically, the series transformation introducing control parameters $u$ can be 
denoted as
\be
\label{3}
 F_k(x,u) = \hat R[\; u \; ] f_k(x) \;  .
\ee
Control functions $u_k(x)$ are to be defined from optimization conditions inducing 
convergence of perturbation theory. Then the sequence of the optimized approximants
\be
\label{4}
\overline f_k(x) = F_k(x,u_k(x))
\ee
is the main result of the optimized perturbation theory \cite{Yukalov_3,Yukalov_4}.
In particular cases, control functions can become control parameters, $u_k(x) = u_k$.

Since control functions have to provide convergence for a sequence of approximants,
the optimization conditions can be derived from the Cauchy criterion of convergence.
The general way for deriving optimization conditions is thoroughly explained in 
review articles \cite{Yukalov_5,Yukalov_6}. The most often used optimization conditions
are the minimal difference condition
\be
\label{5}
F_{k+1}(x,u) - F_k(x,u) = 0   
\ee
and the minimal derivative conditions
\be
\label{6}
\frac{\prt F_k(x,u)}{\prt u}  = 0 \;   .
\ee
When these conditions enjoy unique solutions for the control functions $u_k(x)$ or control
parameters $u_k$, the optimized approximants $(\ref{4})$ provide good accuracy for the
physical quantities of interest \cite{Yukalov_5,Yukalov_6}. However, the problem arises
when the optimization conditions (\ref{5}) or (\ref{6}) do not have unique solutions but 
exhibit multiple solutions, which happens quite often. Then it is not clear which of the
solutions to prefer. The similar problem is typical for the use of the self-similar roots 
\cite{root}, and additive approximants \cite{add}. Strictly speaking, the above 
optimization conditions have to be understood in the general sense of looking for the 
minimal difference in the left-hand side of (\ref{5}) or minimal derivative in the 
left-hand side of (\ref{6}) that do not need to be exactly zero.

\section{Self-Similar Borel-Type Transforms}

Let us specify how control parameters can be introduced by means of sequence 
transformations. For concreteness, let us consider Borel-type transforms. We call
the latter self-similar, since we sum these transforms by employing self-similar root
approximants (see, e.g., \cite{Yukalov_5,Yukalov_6}). The summation procedure 
for the Borel transform $B_k(x,u)$ gives a self-similar iterated root approximant 
\cite{itr} in the form
$$
B_k^*(x,u) = \left( \left( ( 1 + A_1 x)^2 + A_2 x^2 \right)^{3/2} +
\ldots + A_k x^k \right)^{\bt/k} \;  ,
$$    
with all parameters $A_i = A_i(u)$ defined by the accuracy-through order procedure.

\subsection{Self-Similar Borel-Leroy Summation}

The Borel-Leroy transform \cite{Hardy_1,Glimm_7} of the given truncated series (\ref{1}) 
is
\be
\label{7}
B_k(x,u) = \sum_{n=0}^k \frac{a_nx^n}{\Gm(n+u+1)} \;   ,
\ee
where $u$ is a control parameter. Expansion (\ref{7}) can be summed by means of 
self-similar approximants in the form of either factor approximants 
\cite{Gluzman_8,Yukalov_9} or root approximants \cite{Yukalov_10}. If a self-similar 
approximant $B_k^*(x,u)$ is found, then the inverse transform
\be
\label{8}
f_k^*(x) = \int_0^\infty B_k^*(xt,u) t^u e^{-t} \; dt
\ee
gives the self-similar Borel-Leroy summation of the initial series (\ref{1}). 

Finding the large-variable limiting behavior  
\be
\label{9}
 B_k^*(x,u)  \simeq C_k(u) x^\bt \qquad ( x\ra \infty)   
\ee
yields the large-variable behavior for the approximant (\ref{8}),
\be
\label{10}
f_k^*(x) \simeq B_k(u) x^\bt \qquad ( x\ra \infty) \;   ,
\ee
with the critical amplitude
\be
\label{11}
B_k(u) = C_k(u) \Gm(\bt+ u + 1) \;   .
\ee
Then defining the control parameter $u_k$ results in the optimized approximation for
the critical amplitude $B_k \equiv B_k(u_k)$.

\subsection{Self-Similar Mittag-Leffler Summation}

The Mittag-Leffler \cite{Hardy_1,Mittag_11} transform reads as
\be
\label{12}
 B_k(x,u) = \sum_{n=0}^k \frac{a_nx^n}{\Gm(nu+1)} \;  ,
\ee
with $u$ being a control parameter. Constructing a self-similar approximation 
$B_k^*(x,u)$ for (\ref{12}), we come to the self-similar Mittag-Leffler summation
\be
\label{13}
 f_k^*(x) = \int_0^\infty B_k^*(xt^u,u) \; e^{-t} \; dt \;  .
\ee  
The large-variable behavior of the latter approximant has the form (\ref{10}) with the 
amplitude
\be
\label{14}
B_k(u) = C_k(u) \Gm(\bt u +1 ) \;  .
\ee
Defining from optimization conditions control parameters $u_k$, we get the optimized
amplitude $B_k \equiv B_k(u_k)$. 

Note that the formal substitution $u = 1 +v/n$ makes the Mittag-Leffler transform
(\ref{12}) similar to the Borel-Leroy transform (\ref{7}). It is, therefore, reasonable 
to expect that the accuracy of both types of the approximations should be close to each 
other.

\subsection{Use of Fractional Derivatives}
\label{frr}

Borel-type transforms can be introduced employing fractional derivatives, as is mentioned 
in \cite{fr}. The fractional derivative $D^u_x$, with respect to a variable $x$, according 
to the Riemann-Liouville convention \cite{fra1,fra2}, can be applied to power series or 
polynomial expressions according to the prescription
\be
\label{15}
D^u_x x^n=\frac{\Gamma(n+1)}{\Gamma(n-u+1)}x^{n-u}.
\ee

Defining the Borel-type transform
\be
\label{16}
B_k\left(x,u\right) = \sum_{n=0}^k \frac{\Gamma(n-u+1)}{\Gamma^2(n+1)} \; a_n x^n \; ,   
\ee
and constructing a self-similar approximation $B_k^*(x,u)$, we have the inverse 
transformation representing the self-similar approximation
\be
\label{17}
f_k^*(x) = \int_0^\infty e^{-t} (xt)^u D^u_{xt} \left[B_k^*(xt,u)\right] \; dt \;  .
\ee
 
Then the large-variable amplitude of the function (\ref{17}) is 
\be
\label{18}
B_k(u) = C_k(u) \frac{\Gamma^2(\bt+1)}{\Gamma(\bt-u+1)} \; .
\ee

\section{Use of Fractional Integrals}

Let us consider the integral operator, acting on a function $f(t)$ as
\be
\label{19}
I_t f(t) =\frac{1}{t} \int_0^t f(\tau) d\tau \; .
\ee
The repeated application of this integral $u$ times to $t^\bt$ gives
\be
\label{20}
I_t^u t^\bt = \frac{1}{(\bt + 1)^u} t^\bt \; .
\ee
Extending this to any noninteger $u$ defines a fractional integral.

It is possible to define \cite{fr} the Borel-type transform 
\be
\label{21}
B_k(x,u) = \sum_{n=0}^k \frac{(n+1)^u}{\Gamma(n+1)} \; a^n x^n \; ,
\ee
where the parameter $u$ can be arbitrary. This formula is well-defined for $n=0$.

When the transformed series (\ref{21}) is summed by means of a self-similar approximation
yielding $B_k^*(x,u)$, then the sought function is approximated by the expression
\be
\label{22}
f_k^*(x) = \int_0^\infty e^{-t} I_t^u B_k^*(xt,u) \; dt \; .
\ee
The large-variable amplitude of the sought function (\ref{22}) reads as
\be
\label{23}
B_k(u) = C_k(u) \frac{\Gm(\bt+1)}{(\bt+1)^u} \; .
\ee

\section{Lasso and Ridge Selection Criteria}

In the previous section, the ways of introducing control parameters $u$ by means of 
Borel-type transforms are presented. When the control parameters are defined with the 
optimization conditions, discussed in Sec. 2, often the problem of multiplicity of 
solutions appears. There are two approaches to consider the problem of solution 
non-uniqueness. First approach suggests to consider all possible solutions to the 
optimization problem and select the unique representative according to some natural, 
plausible selection criteria. Second approach requires such a definition of the 
optimization problem so that the solution is always unique. 

Let us proceed with the first approach and consider the optimization conditions with 
respect to the sought quantity that is the large-variable amplitude $B_k(u)$. We have 
two optimization conditions, the minimal difference condition,
\be
\label{24}
B_{k+1}(u) - B_k(u) = 0 
\ee
and the minimal derivative condition,
\be
\label{25}
 \frac{d B_k(u)}{d u} = 0 \;  .
\ee
Solving separately both these equations, we can obtain several solutions for the control 
parameters, all of which we enumerate as $u_j$, with $j = 0,1,2,\ldots$, where $u_0 =0$ 
for all Borel-type transforms, except the Mittag-Leffler transform, for which $u_0=1$.   
These optimization conditions are equivalent \cite{Yukalov_5,Yukalov_6} and it should 
be quite natural to consider the solutions to optimizations as equivalent as well and 
treat them all together.

All Borel-type transforms can be represented as the sums
\be
\label{26}
B_k(x,u) = \sum_{n=0}^k b_n(u) x^n \;   ,
\ee
with the appropriate coefficients. Studying the coefficients $b_n(u_j)$ it is possible 
to introduce several measures for selecting an optimal control parameter $u_{opt}$.

Lasso measures were introduced for multivariate regression construction to be obtained 
under the additional condition of minimal sums of the regression coefficients 
\cite{las1, las2}. Motivated by the lasso criteria, it is reasonable to impose a selection 
criteria, so that the solution to optimization problems would correspond to the minimal 
sum of the transformed coefficients. Thus we need to find the values of the lasso 
functional
\be
\label{27}
L_1(u_j) = \sum_{n=0}^k |\; b_n(u_j) \; | \;   .
\ee
The solution $u_{opt}$ corresponding to the minimal lasso functional, such that
\be
\label{28}
L_1(u_{opt}) =  \min_j \; L_1(u_j) \; ,
\ee
gives the optimal control parameter. 

The close condition is the minimization of the sum of the relative values of the 
transformed coefficients, corresponding to the lasso functional
\be
\label{29}
L_2(u_j) = \sum_{n=0}^k \left|\; \frac{b_n(u_j)}{a_n} \; \right| \; .
\ee
Then the optimal control parameter is given by the minimal $L_2(u_j)$. 

One can require that the solution to be chosen should correspond to the optimization 
with the minimal possible perturbation of the original truncation. Alternatively, one 
can think that the chosen solution should minimally perturb the Borel-transformed 
coefficients, since we expect that the Borel transformation and summation should be 
effective and accurate by its own. Such approaches are within the scope of classical 
Tikhonov regularization and generalized Tikhonov regularization \cite{t1,t2}, the 
techniques devised specifically to untangle the problem of multiple solutions.

In the spirit of the generalized Tikhonov regularization \cite{t1,t2}, let us define 
the generalization of the lasso selection criteria requiring the preferred solution 
to deviate minimally from the solution with Borel-transformed coefficients. Thus, 
the functional 
\be
\label{30}
G_1(u_j) = \frac{1}{k+1}\sum_{n=0}^k 
\left|\; \frac{b_n(u_j)-b_n(u_0)}{b_n(u_0)} \; \right| 
\ee
defines how strongly the transformed coefficients with control parameters deviate from 
the Borel-transformed coefficients without such parameters. 

To compute $G_1(u_j)$, one should solve the optimization problem consisting of finding 
all solutions to the minimal derivative and minimal difference optimization conditions 
and then pick the solution with $u_j$ minimizing functional (\ref{30}).

The other generalization of the lasso selection criteria, requires that the preferred 
solution would minimally deviate from the solution with non-transformed coefficients.
The functional
\be
\label{31}
G_2(u_j) = \frac{1}{k+1}\sum_{n=0}^k 
\left|\; \frac{b_n(u_j)-a_n}{a_n} \; \right| \;
\ee
defines how strongly the transformed coefficients deviate from the original, 
non-transformed coefficients. Here again, we should solve the optimization problem 
consisting of finding all solutions to the minimal derivative and minimal difference 
conditions, and then pick the solution with $u_j$ minimizing functional (\ref{31}).

In the other approach, it is necessary to construct a cost functional to be directly
minimized with respect to the control parameter considered as a variable. Then it is
possible to introduce a functional that is a combination of the minimal-difference and 
minimal-derivative conditions, plus a ridge-type penalty term introduced to select the 
unique solution deviating minimally from the pure Borel solution. Such a cost functional 
with a ridge penalty is 
\be
\label{32}
F(u)=\lambda |B_{k+1}(u)-B_{k}(u)|^2+(1-\lambda ) \left|\frac{d {B_{k}}(u)}{du}\right|^2
+\frac{1}{2} (u-u_0)^2 \; ,
\ee
with $\lambda =1/2$ and $u_0=0, 1$ for the Borel solution, depending on the type 
of a transformation. 

The optimal control parameter is that one minimizing the cost functional (\ref{32}), 
so that
\be
\label{33}
 F(u_{opt}) = \min_u F(u) \;  .
\ee

\section{Comparison of Methods}

In the present section, we consider several typical problems in order to illustrate the 
applicability of the suggested methods and to compare the accuracy of different approaches.  
In the tables and in the text below, for the sake of simplicity, we call the method based
on the functional $L_1(u_j)$, lasso$_1$, based on functional $L_2(u_j)$, lasso$_2$, based 
on functional $G_1(u_j)$, genlass$_1$, based of $G_2(u_j)$, genlass$_2$, and the method 
using the minimization of $F(u)$, just the functional method.  

We bring up only the results obtained in the highest possible order for each of the 
problem, limited only by the number of terms in the perturbative expansions, or by the 
highest possible order, where analytical calculations can be performed with MATHEMATICA.
Complex valued solutions are discarded. As is natural, the results in lower orders are 
worse than in the higher orders presented below.

\subsection{Schwinger Model: Energy Gap}
\label{GAP}

The Schwinger model is considered as a touchstone in high energy physics. The spectrum 
of bound states for the Schwinger model can be presented in the form of asymptotic 
expansions \cite{Hamer_16}. Thus the energy gap $\Delta(z)$ for the so-called scalar 
state at small $z = (1/ga)^4$, where $g$ is a coupling parameter and $a$ is the lattice 
spacing, can be represented as a truncated series 
\be
2 \Dlt(z) \simeq
1+6 z-26 z^2+190.6666666667 z^3-1756.666666667 z^4+18048.33650794 z^5 \; .
\ee
The coefficients $a_n$ are rapidly increasing by absolute value. They are known up to 
the $13$-th order. In the continuous limit of the lattice spacing tending to zero the gap 
acquires the form of a power-law \cite{Hamer_16},
\be
\label{04}
\Dlt(z) \simeq  B z^{\bt} \qquad ( z \ra \infty),
\ee
where $B=1.1284$, $\bt=1/4$. The results of calculations by different methods are presented 
in Table \ref{Table 1}. The most accurate results are given in bold.

To illustrate the case, we present the results obtained from the fractional integration. 
There are altogether three solutions to the optimization problems. Equation \eqref{5} 
possesses a single solution, $$B_{1}=1.35252,$$ corresponding to $u_{1}=-5.46262.$  
Strictly speaking, in such a case the optimal value $u_{1}$ 
corresponds to the minimal difference between the terms in the left-hand-side  
of the \eqref{5}, but uncertainty in the results, $B_{1}=1.35252 \pm 0.00056$, 
caused by such circumstances is very small.

Equation \eqref{6} possesses two solutions,
$$B_{2}=1.35544,\,\,\,\,\,B_{3}=1.36432,$$ corresponding to 
$$u_{2}=-5.641492,\,\,\,\,\,u_{3}=-5.873576.$$

\begin{table}
\caption{Energy Gap of the Schwinger Model: Amplitude ($k=11$)
}
\label{Table 1}       
\centering
\begin{tabular}{|c|c|c|c|c|c|c|c|c|c}
\hline
$B$ & genlass$_{1}$ & genlass$_{2}$ & lasso$_{1}$ & lasso$_2$ &functional \\\hline
Mittag-Leffler&1.20788 &1.20788&1.20788 &1.20788 &1.17234\\\hline
Borel-Leroy &--&--&--  &-- &1.16805\\ \hline
Fract. Riemann &1.55562&1.55562&1.58458&1.55562&\bf{1.15757}\\\hline
Fract. Integral&1.35252&1.35252 &1.36432&1.36432&1.16269\\\hline
\end{tabular}
\end{table}

Borel-Leroy summation with optimization conditions \eqref{5}, \eqref{6} fails to produce 
real solutions. Good results are achieved by the functional method, due to the good 
performance of the ``pure'' Borel summation technique \cite{bo}. The best estimate for 
the amplitude in the $11$th order of perturbation theory, $B=1.11717$. 

For the so-called vector or ground sate of the Schwinger model, the best estimate for 
the amplitude could be obtained in the $11$th order of perturbation theory, $B=0.566622$, 
by applying the minimal derivative method of Fractional Borel summation developed in the 
paper \cite{fr}. The latter estimate is very close to  the exact result for the critical 
amplitude, $B=0.5642$ \cite{Hamer_16}. In the $10$-th order already quite good results
$B=0.579718$, achieved by the minimal derivative optimization, are presented and discussed 
in the paper \cite{fr}. Close results, $B=0.579717$, can be found from the functional 
method in the $11$th order. All other methods discussed above in the best case give 
$B\approx 0.63$. The problems is extremely difficult because neither the original 
iterated roots or pure Borel summation manage to produce real numbers in high-orders of 
perturbation theory. The series for vector and scalar states are notoriously difficult 
for resummation by major techniques, as discussed in the paper \cite{Hamer_16}.

\subsection{Schwinger Model: Critical Amplitude}
\label{Sch2}

The ground-state energy $E$ of the Schwinger  model at small-$x$ can be expanded as 
follows, \cite{Hamer_16,Carrol_29,Vary_30,Adam_31,Striganesh_32,Coleman_33,Hamer_34}:
\begin{equation}
\label{37}
 E(x) \simeq 0.5642 - 0.219 x + 0.1907 x^2 \qquad ( x \rightarrow 0 ) \; .
\end{equation}
The large-$x$ behavior is known as well: 
$$
E(x) \simeq B x^{\bt} \qquad ( x \rightarrow \infty ),
$$
behaving as a power-law, with $B=0.6418$, $\bt=-1/3$.
 
The results of calculations by different methods are shown in Table \ref{Table 2}. The 
most accurate results are given in bold.

\begin{table}
\caption{Schwinger Model. Second Order : Amplitude ($k=2$)
}
\label{Table 2}       
\centering
\begin{tabular}{|c|c|c|c|c|c|c|c|c|c}
\hline
$B$ & genlass$_{1}$ & genlass$_{2}$ & lasso$_{1}$ & lasso$_2$ &functional \\\hline
Mittag-Leffler&0.551877 & 0.551877&0.551877 &0.551877 & 0.62336\\\hline
Borel-Leroy &0.551495 &0.551495  &0.550929 &0.550929& \bf{0.647848}\\ \hline
Fract. Riemann &--& --&--&--&0.696407\\\hline
Fract. Integral&0.51141&0.51141 &0.473248&0.473248&0.672065\\\hline
\end{tabular}
\end{table}

The results of Borel-Leroy summation with functional integration applied to find the 
control parameters, are very good and are in accordance with the best results shown 
by different methods \cite{fr}. The method of fractional derivatives of section \ref{frr} 
produces zero amplitude. However the functional method gives quite good results.

Let us briefly comment on the subject of improving the performance of the fractional 
integration summation. Sometimes, a naive addition of one more trial term with $a_3=0$ 
may help to improve the results. For instance, application of the method of fractional 
integration in the third order accomplished along the same lines as above, gives much 
better results. They are presented in Table \ref{Table 3}. The most accurate results 
are given in bold.

To illustrate the case of an amplitude at infinity of the Schwinger model we present 
the results obtained from the fractional integration. In the case of amplitude of the 
Schwinger model there are two solutions to the optimization problems. Equation \eqref{5} 
possesses a single  solution,
$$B_{1}=0.615054,$$ corresponding to $u_{1}=-0.60088$.
Equation \eqref{6} also possesses a single  solution,
$$B_{2}=0.59844,$$ corresponding to $u_{2}=-1.11869$.

Strength of the functional 
estimates in such cases is granted by a very good performance of the ``pure'' Borel 
method. It produces the result $B=0.6562$, even without any optimization.
\begin{table}
\caption{Schwinger Model. Third Order: Amplitude ($k=3$)
}
\label{Table 3}       
\centering
\begin{tabular}{|c|c|c|c|c|c|c|c|c|c}
\hline
$B$ & genlass$_{1}$ & genlass$_{2}$ & lasso$_{1}$ & lasso$_2$ &functional \\\hline
Fract. Integral&0.615054&0.615054 &0.59844&0.59884&0.65596\\\hline
\end{tabular}
\end{table}

\subsection{Anomalous Dimension}
\label{ANOM}

In the $n=4$ supersymmetric Yang-Mills theory, the cusp anomalous dimension $\Omega(g)$ 
of a light-like Wilson loop depends only on the coupling $g$ (see \cite{Banks,YGh} and 
references therein).

In terms of the variable $x=g^2$, the problem can be written down in terms of the function  
$f(x)=\frac{\Omega(x)}{x}$, with the following weak-coupling expansion,
$$ 
f(x) \simeq 4-13.1595 x+95.2444 x^2-937.431 x^3,  \quad x \rightarrow 0,
$$
while in the strong-coupling limit $f(x)$ takes the form of a power-law 
$$
f(x) \simeq B x^{\bt}, \quad x \rightarrow \infty \; ,  
$$
with $B=2$ and $\beta=-1/2$.

The results of calculations by different methods are shown in Table \ref{Table 4}. The 
most accurate results are given in bold.

To illustrate the case of an anomalous dimension we present the results obtained from 
the fractional integration. In the case of anomalous dimension there are altogether 
four solutions to the optimization problems.  Equation \eqref{5} possesses two  solutions,
$$B_{1}=2.0488,\,\,\,\,B_{2}=1.70916,$$ corresponding to 
$$u_{1}=-0.780063,\,\,\,\,u_{2}=-2.575136.$$

Equation \eqref{6} also possesses two  solutions,
$$B_{3}=1.8758,\,\,\,\,B_{4}=2.22766,$$ corresponding to 
$$u_{3}=-1.557644,\,\,\,\,u_{4}=-2.29044.$$
\begin{table}
\caption{Anomalous Dimension: Amplitude  ($k=3$)
}
\label{Table 4}       
\centering
\begin{tabular}{|c|c|c|c|c|c|c|c|c|c}
\hline
$B$ & genlass$_{1}$ & genlass$_{2}$ & lasso$_{1}$ & lasso$_2$ &functional \\\hline
Mittag-Leffler&2.08308 & 1.93162&2.08308 &2.08308 & \bf{1.99381}\\\hline
Borel-Leroy &2.01177 &2.01177  &1.93142 &1.93142& 2.05526\\ \hline
Fract. Riemann &1.52111&1.52111&1.52111&1.52111&2.44164\\\hline
Fract. Integral&2.0488&2.0488 &1.70916&1.70916&2.32582\\\hline
\end{tabular}
\end{table}

Performance of the functional method in the case of Mittag-Leffler summation is exemplary, 
since neither the original iterated roots or pure Borel summation work well enough by 
themselves. Other criteria also work quite well.

\subsection{Quantum Quartic Oscillator}
\label{OSC}
The anharmonic oscillator is described by the model Hamiltonian in which non-linearity 
is quantified through a positive coupling (anharmonicity) parameter $g$ \cite{Hio1978}. 
Perturbation theory for the ground-state energy yields \cite{Hio1978} a rather long 
expansion 
$
 E_k(g) = \sum_{n=0}^k a_n g^n .
$
The starting coefficients are
$$
a_0 = \frac{1}{2} \; , \qquad a_1 = \frac{3}{4} \; , \qquad
a_2 = -\; \frac{21}{8} \;  \qquad a_3 = \frac{333}{16} \; , \qquad
a_4 = -\; \frac{30885}{128} \; .
$$
The coefficients $a_n$ rapidly grow in magnitude.

The strong-coupling limit-case, as $g \ra \infty$, 
\be
\label{6.7}
 E(g) \simeq B g^{\beta}  \; ,
\ee
is known. The parameters $B=0.667986,$ $\beta=1/3$, are known as well. The results 
of calculations by different methods are shown in Table \ref{Table 5}. The most accurate 
results are given in bold.

To illustrate the case of a quantum quartic oscillator we present the results obtained from the fractional integration. In the case of quantum quartic oscillator there are altogether eight solutions to the optimization problems.  Equation \eqref{5} possesses four  solutions,
$$B_{1}=0.676006,\,\,\,\,B_{2}=0.681481,\,\,\,\,B_{3}=0.671669,\,\,\,\,B_{4}=0.694167,$$ corresponding to 
$$u_{1}=-0.303653,\,\,\,\,u_{2}=-0.668542,\,\,\,\,u_{3}=-1.094372,\,\,\,\,u_{4}=-1.448501.$$

Equation \eqref{6} also possesses four  solutions,
$$B_{5}=0.67498,\,\,\,\,B_{6}=0.681551,\,\,\,\,B_{7}=0.669679,\,\,\,\,B_{8}=0.69537,$$  
$$u_{5}=-0.179503,\,\,\,\,u_{6}=-0.641464,\,\,\,\,u_{7}=-1.029708,\,\,\,\,u_{8}=-1.40472.$$
\begin{table}
\caption{Quantum Quartic Oscillator : Amplitude  ($k=11$)
}
\label{Table 5}       
\centering
\begin{tabular}{|c|c|c|c|c|c|c|c|c|c}
\hline
$B$ & genlass$_{1}$ & genlass$_{2}$ & lasso$_{1}$ & lasso$_2$ &functional \\\hline
Mittag-Leffler&0.682136 & 0.682136&0.692631 &0.692631 & 0.679929\\\hline
Borel-Leroy &0.674414 &0.674414  &0.677297 &0.677297& 0.677097\\ \hline
Fract. Riemann &\bf{0.674121}&\bf{0.674121}&0.689068&0.674477&0.677112\\\hline
Fract. Integral&0.67498&0.67498 &0.694167&0.694167&0.679967\\\hline
\end{tabular}
\end{table}

Different ``genlass'' measures perform better than ``lasso'' measures. The results 
remain close to the numbers $B=0.676708$, given by the pure Borel summation. 

The original iterated roots fail in the higher-orders producing complex numbers. Yet, 
the best estimate for the amplitude in the $11$th order of optimized perturbation theory, 
$B=0.667638$, can be found by applying the minimal difference optimization of the 
Fractional Borel summation suggested in the paper \cite{fr}. In the $10$-th order, 
already quite good results, $B=0.669356$, are obtained \cite{fr}.

\subsection{Three-dimensional Harmonic Trap}
\label{Trap}
 
Quantum properties of Bose-condensed atoms in a spherically-symmetric harmonic trap 
can be found from the three-dimensional stationary nonlinear Schr\"{o}dinger equation 
\cite{BCY}. The problem can be reduced to studying only the radial part of the condensate 
wave function. The ground state energy $E$ of the trapped Bose-condensate can be 
approximated by the following truncation
\begin{equation}
\label{spw}
E(c)\simeq \frac{3}{2}+\frac{1}{2} c -\frac{3}{16} c^2+\frac{9}{64} c^3 -\frac{35}{256} c^4 
\qquad(c\rightarrow 0) \; ,
\end{equation}
where $c$ plays the role of an effective coupling parameter taking account of trapping. 

For very strong trapping, the energy behaves as the power-law
\begin{equation}
\label{sps}
E(c)\simeq B c ^{\bt} 
\qquad (c\rightarrow \infty),
\end{equation}
with the amplitude at infinity $B=5/4$, and the index $\bt=2/5$ \cite{BCY}.

The results of calculations by different methods are shown in Table \ref{Table 6}. The 
most accurate results are given in bold.

To illustrate the case of a three-dimensional harmonic trap we present the results 
obtained from the fractional integration. In the case of three-dimensional harmonic 
trap there are altogether six solutions to the optimization problems. Equation \eqref{5} 
possesses three  solutions,
$$B_{1}=1.28602,\,\,\,\,B_{2}=1.28725,\,\,\,\,B_{3}=1.32299,$$ corresponding to 
$$u_{1}=-0.101644,\,\,\,\,u_{2}=-2.352295,\,\,\,\,u_{3}=-3.098849.$$

Equation \eqref{6} also possesses three  solutions,
$$B_{4}=1.28782,\,\,\,\,B_{5}=1.27625,\,\,\,\,B_{6}=1.32334,$$  
$$u_{4}=-0.478447,\,\,\,\,u_{5}=-1.80111,\,\,\,\,u_{6}=-3.14702.$$

\begin{table}
\caption{Quantum Trap: Amplitude  ($k=4$)
}
\label{Table 6}       
\centering
\begin{tabular}{|c|c|c|c|c|c|c|c|c|c}
\hline
$B$ & genlass$_{1}$ & genlass$_{2}$ & lasso$_{1}$ & lasso$_2$ &functional \\\hline
Mittag-Leffler&1.28579 & \bf{1.28162}&1.28579 &1.28579 & 1.28553\\\hline
Borel-Leroy &1.28664&1.28664  &1.28664 &1.28951& 1.28523\\ \hline
Fract. Riemann &1.28677&1.28677&1.28677&1.28677&1.28473\\\hline
Fract. Integral&1.28602&1.28602 &1.32334&1.32334&1.28493\\\hline
\end{tabular}
\end{table}
The results from different methods appear to be very close between each other and with 
other techniques \cite{fr}. Fractional Borel Summation gives $B\approx 1.285$ \cite{fr}, 
in line with the estimates from Table \ref{Table 6}. The most accurate results are given 
in bold.

\subsection{One-Dimensional Quantum Nonlinear Model}
\label{nonl}

The energy levels $E$ of Bose-condensed interacting atoms in a one-dimensional harmonic 
trap can be represented \cite{BCY} by the form
$$
E(g) = \left ( n + \frac{1}{2} \right ) f(g) \; ,
$$
where $n = 0,1,2,\ldots$ is a quantum index and $g$ is a dimensionless coupling parameter 
quantifying the effect of trapping. Perturbation theory for the function $f(g)$ gives the 
expansion
$$
f_k(g) = 1 + \sum_{n=1}^k a_n z^n  
$$
in powers of $g$. For $k=5$ the coefficients $a_n$ can be found in the paper \cite{BCY}:
$$
a_1 = 1 , \, \,\,a_2 = - \; \frac{1}{8} ,\, \,\,
a_3 = \frac{1}{32} ,\,\, \,  a_4 = - \; \frac{1}{128},\, \, \,a_5 =\frac{3}{2048} \; .
$$

The strong-coupling limit in the leading order is of power-law,
$$
f(g) \simeq B g^{\beta} \; ,
$$
with $B=3/2$, $\beta=2/3$.

The results of calculations by different methods are shown in Table \ref{Table 7}. The
Borel-Leroy summation with optimization conditions \eqref{5}, \eqref{6} do not enjoy
real solutions. 

To illustrate the case of a one-dimensional quantum nonlinear model we present the 
results obtained from the fractional integration. In the case of one-dimensional 
quantum nonlinear model there are altogether four solutions to the optimization 
problems.  Equation \eqref{5} possesses two  solutions,
$$B_{1}=1.48168,\,\,\,\,B_{2}=1.50372,$$ corresponding to 
$$u_{1}=-4.268415,\,\,\,\,u_{2}=-5.979522.$$

Equation \eqref{6} also possesses two  solutions,
$$B_{3}=1.4809,\,\,\,\,B_{4}=1.50378,$$ corresponding to 
$$u_{3}=-4.030307,\,\,\,\,u_{4}=-6.016694.$$

\begin{table}
\caption{Quantum One-Dimensional Trap: Amplitude  ($k=5$)
}
\label{Table 7}       
\centering
\begin{tabular}{|c|c|c|c|c|c|c|c|c|c}
\hline
$B$ & genlass$_{1}$ & genlass$_{2}$ & lasso$_{1}$ & lasso$_2$ &functional \\\hline
Mittag-Leffler&1.44736 & 1.44795&1.44736 &1.44736 & 1.36429\\\hline
Borel-Leroy &--&--  &-- &--& 1.38647\\ \hline
Fract. Riemann &1.53989&1.53989&1.53989&1.53989&1.38512\\\hline
Fract. Integral&1.4809&1.4809 &\bf{1.50378}&\bf{1.50378}&1.36135\\\hline
\end{tabular}
\end{table}
Functional methods which are based on the strength of pure Borel method are inferior to 
other approaches. Obviously, Borel summation is not so good in the case of a $1d$ nonlinear 
model. Fractional integral combined with lasso selection criteria works very well, 
even better than $B \approx 1.475$, obtained from the Fractional Borel summation \cite{fr}.

\subsection{Two-Dimensional Polymer: Swelling}
\label{2dp}

For the swelling factor $\Upsilon$ of the two-dimensional polymer, perturbation theory 
developed in \cite{Muthukumar_1} yields  the expansion in powers of the dimensionless 
coupling parameter $g$. The parameter quantifies the repulsive interaction between the  
segments of the polymer chain \cite{Muthukumar_1,Muthukumar_2}. As $g\rightarrow 0$ the 
swelling factor can be represented as the following truncation,
\begin{equation}
\label{633}
\Upsilon(g) \simeq 1 + \frac{1}{2} \; g - 0.12154525 \; g^2+ 0.02663136 \,g^3-0.13223603\,g^4.
\end{equation}
As $g\rightarrow \infty$, the swelling factor behaves as a power-law, i.e., 
$\Upsilon(g)\simeq B g^\bt.$ The index at infinity $\bt=1/2$ is known exactly ~\cite{grosb,
pel}. The amplitude $B$ is of the order of unity. The results of calculations by different 
methods are shown in Table \ref{Table 8}. They all give very close numbers which appear 
to be close to unity. Fractional Borel Summation gives $B\approx 0.971$ \cite{fr}, in 
line with other methods.

To illustrate the case of a two-dimensional polymer, we present the results obtained 
from the fractional integration. In the case of two-dimensional polymer, there are two 
solutions to the optimization problems. Equation \eqref{5} possesses a single  solution,
$$B_{1}=0.974145,$$ corresponding to $u_{1}=-0.401295$.
Equation \eqref{6} also possesses a single  solution,
$$B_{2}=0.974499,$$ corresponding to $u_{2}=-0.533987$.

\begin{table}
\caption{Two-Dimensional Polymer: Amplitude  ($k=4$)
}
\label{Table 8}       
\centering
\begin{tabular}{|c|c|c|c|c|c|c|c|c|c}
\hline
$B$ & genlass$_{1}$ & genlass$_{2}$ & lasso$_{1}$ & lasso$_2$ &functional \\\hline
Mittag-Leffler&0.972576 & 0.972582&0.972576 &0.972576 & 0.9703\\\hline
Borel-Leroy &0.975689&0.975689  &0.976097 &0.976097& 0.969957\\ \hline
Fract. Riemann &\bf{0.97779}&\bf{0.97779}&\bf{0.97779}&\bf{0.97779}&0.969277\\\hline
Fract. Integral&0.974145&0.974145 &0.974499&0.974499&0.969564\\\hline
\end{tabular}
\end{table}

\subsection{Three-Dimensional Polymer: Amplitude}
\label{3dp}

Similarly, the expansion factor of a three-dimensional polymer can be represented as 
a truncated series in a single dimensionless interaction parameter $g$ 
\cite{Muthukumar_1,Muthukumar_2}. As $g\ra 0$, the expansion factor $\Upsilon(g)$ can 
be presented as the truncated series with the coefficients
$$
a_0 = 1 \; , \qquad a_1 = \frac{4}{3} \; , \qquad a_2 = -2.075385396 \; \qquad
a_3 = 6.296879676 \; ,
$$
$$
a_4 = -25.05725072,\;, \qquad a_5 = 116.134785 \; ,  \qquad
a_6 = -594.71663 \; .
$$
The strong-coupling behaviour of the expansion factor, as $g\ra \infty$, is of power-law,
$\Upsilon(g)\simeq B g^\bt.$ It has been found numerically \cite{Muthukumar_2}, that 
$B\approx 1.5309$, while $\bt\approx 0.3544$.

The results of calculations by different methods are shown in Table \ref{Table 9}. They 
all produce rather close numbers which appear to be close to numerical estimates. 
Fractional Borel Summation gives $B\approx 1.535$ \cite{fr}, in line with other  
results presented in Table \ref{Table 9}.

To illustrate the case of a three-dimensional polymer we present the results obtained 
from the fractional integration. In the case of three-dimensional polymer  there are 
altogether six solutions to the optimization problem.  Equation \eqref{5} possesses 
three  solutions,
$$B_{1}=1.53565,\,\,\,\,B_{2}=1.54271,\,\,\,\,B_{3}=1.53362,$$ corresponding to 
$$u_{1}=-0.73769,\,\,\,\,u_{2}=-1.86859,\,\,\,\,u_{3}=-3.14257.$$

Equation \eqref{6} also possesses three  solutions,
$$B_{4}=1.53685,\,\,\,\,B_{5}=1.54634,\,\,\,\,B_{6}=1.53082,$$ corresponding to 
$$u_{4}=-1.040533,\,\,\,\,u_{5}=-2.31288,\,\,\,\,u_{6}=-3.03694.$$

\begin{table}
\caption{Three-Dimensional Polymer: Amplitude  ($k=6$)
}
\label{Table 9}       
\centering
\begin{tabular}{|c|c|c|c|c|c|c|c|c|c}
\hline
$B$ & genlass$_{1}$ & genlass$_{2}$ & lasso$_{1}$ & lasso$_2$ &functional \\\hline
Mittag-Leffler&1.53574 & 1.53765&1.53574 &1.53574 & 1.52826\\\hline
Borel-Leroy &\bf{1.53228}&\bf{1.53228}  &1.53267 &1.53267& 1.52607\\ \hline
Fract. Riemann &1.54154&1.54154&1.56952&1.54154&1.52693\\\hline
Fract. Integral&1.53565&1.53565 &1.53362&1.53362&1.52728\\\hline
\end{tabular}
\end{table}

\subsection{Bose Temperature Shift}
\label{BS}

The Bose-Einstein condensation temperature $T_0$ of the ideal uniform Bose gas in a
three-dimensional space is well-known. However, the ideal Bose gas is unstable below 
the condensation temperature \cite{BCY}. Atomic interactions stabilize the system and 
induce the shift $\Dlt T_c \equiv T_c - T_0$ of the Bose-Einstein condensation 
temperature $T_c$ of a non-ideal Bose system (see review \cite{YY}). 

At asymptotically small gas parameter
$
\gm \equiv \rho^{1/3} a_s \; ,
$
where $a_s$ is atomic scattering length and $\rho$ stands for the gas density, the shift 
is quantified by the parameter $c_1$, i.e.,
$$
 \frac{\Dlt T_c}{T_0} \simeq c_1 \gm \; ,
$$ 
for $\gm \ra 0 $.

In order to calculate $c_1$ theoretically, it was suggested to calculate at first an 
auxiliary function $c_1\left(g\right)$ \cite{Kas2004a, Kas2004b, Kas2004c}. Then one 
can find  $c_1$ as the limit 
\be
\label{7.10}
c_1 = \lim_{g\ra\infty} c_1\left(g\right) \equiv B 
\ee
of the function  $c_1\left(g\right)$. The latter limit ought to be found from the 
expansion over an effective coupling parameter,
\be
\label{7.11}
c_1\left(g\right) \simeq a_1 g + a_2 g^2 + a_3 g^3 + a_4 g^4 + a_5 g^5 \;  ,
\ee
where the coefficients 
$$
a_1=0.223286\; , \qquad a_2=-0.0661032 \; , \qquad
a_3=0.026446 \; , $$
$$
a_4=-0.0129177 \; , \qquad a_5=0.00729073 \;  ,
$$
are known and are decreasing monotonically by their absolute values.

Monte Carlo simulations \cite{Arn2001a, Arn2001b, Nho2004}, evaluate
\be
\label{7.9}
c_1 = 1.3. \pm 0.05 \;  .
\ee
The results of calculations by different methods are shown in Table \ref{Table 10}. 

To illustrate the case of a Bose temperature shift, we present the results obtained 
from the fractional integration. In the case of Bose temperature shift there are 
altogether three solutions to the optimization problems.  Equation \eqref{5} possesses 
two solutions,
$$B_{1}=1.26409,\,\,\,\,\,B_{2}=1.05911,$$ corresponding to 
$$u_{1}=-0.76233,\,\,\,\,\,u_{2}=-2.047.$$

Equation \eqref{6} possesses a single solution,
$$B_{3}=1.05429,$$
corresponding to 
$u_{3}=-1.9122.$

\begin{table}
\caption{Bose-Condensation Temperature: Amplitude  ($k=4$)
}
\label{Table 10}       
\centering
\begin{tabular}{|c|c|c|c|c|c|c|c|c|c}
\hline
$B$ & genlass$_{1}$ & genlass$_{2}$ & lasso$_{1}$ & lasso$_2$ &functional \\\hline
Mittag-Leffler&1.33967 & 1.23142&1.33967 &1.33967 & 1.244\\\hline
Borel-Leroy &1.28676&1.28676  &1.18035 &1.18035& 1.37587\\ \hline
Fract. Riemann &\bf{1.28951}&\bf{1.28951}&\bf{1.28951}&\bf{1.28951}&1.53199\\\hline
Fract. Integral&1.26409&1.26409 &1.05911&1.05911&1.54664\\\hline
\end{tabular}
\end{table}

In the same way, one can find the values of $c_1$ for the $O(1)$ field theory with the 
following, formally obtained, expansion available,
$$
c_1(g) \simeq 0.334931 g-0.178478 g^2+0.129786 g^3-0.115999 g^4+0.120433 g^5 \;  ,
$$
with the Monte Carlo numerical estimate $c_1 = 1.09. \pm 0.09$ .

The results of calculations by different methods are shown in Table \ref{Table 11}.

\begin{table}
\caption{Condensation Temperature for the $O(1)$ field theory: Amplitude  ($k=4$)
}
\label{Table 11}       
\centering
\begin{tabular}{|c|c|c|c|c|c|c|c|c|c}
\hline
$B$ & genlass$_{1}$ & genlass$_{2}$ & lasso$_{1}$ & lasso$_2$ &functional \\\hline
Mittag-Leffler&1.14124 & 1.04749&1.14124 &1.04749 & 1.05845\\\hline
Borel-Leroy &\bf{}1.09556&\bf{1.09556}  &0.994172 &0.994172& 1.18093\\ \hline
Fract. Riemann &1.09922&1.09922&1.09922&1.09922&1.30128\\\hline
Fract. Integral&1.07384&1.07384 &1.07384&1.07384&1.31773\\\hline
\end{tabular}
\end{table}

For the $O(4)$ field theory the following expansion can be found,
$$ 
c_1(g) \simeq 0.167465 g-0.0297465 g^2+0.00700448 g^3-0.00198926 g^4+0.000647007 g^5 \; ,
$$
while the Monte Carlo numerical estimate is $c_1 = 1.6. \pm 0.1$. The results of 
calculations by different methods are shown in Table \ref{Table 12}. 
\begin{table}
\caption{Condensation Temperature for the $O(4)$ field theory: Amplitude  ($k=4$)
}
\label{Table 12}       
\centering
\begin{tabular}{|c|c|c|c|c|c|c|c|c|c}
\hline
$B$ & genlass$_{1}$ & genlass$_{2}$ & lasso$_{1}$ & lasso$_2$ &functional \\\hline
Mittag-Leffler&\bf{1.60226} & 1.48142&\bf{1.60226} &\bf{1.60226} & 1.49524\\\hline
Borel-Leroy &1.53953&1.53953  &1.42394 &1.42394& 1.64361\\ \hline
Fract. Riemann &1.54101&1.54101&1.54101&1.54101&1.75795\\\hline
Fract. Integral&1.50931&1.50931 &1.30641&1.30641&1.34281\\\hline
\end{tabular}
\end{table}

Selection with the criteria genlass$_1$ gives the most consistent results in all three cases.

\subsection{One-Dimensional Bose Gas: Amplitude}
\label{BOS}

The ground-state energy $E(g)$ of the one-dimensional Bose gas with contact interactions 
depends only on the dimensionless coupling parameter $g$. The expansion for the energy was 
found in high orders of perturbation theory \cite{Lie1963,ris}. In the variables
$$
e(x)  \equiv E\left ( x^2 \right )\; ,\qquad g \equiv x^2 \; ,
$$ 
the weak-coupling expansion for $x\rightarrow 0$ is
\begin{equation}
\begin{array}{llllllll}
\label{77.4}
e(x) \simeq x^2~ (1 -0.4244131815783876~x+0.06534548302432888~x^2-\\0.001587699865505945~ x^3-0.00016846018782773904~ x^4-\\0.00002086497335840174~x^5-3.1632142185373668 ~10^{-6}~ x^6-\\6.106860595675022 ~10^{-7}~ x^7-1.4840346726187777~ 10^{-7}~ x^8) \; .
\end{array}
\end{equation}
In the limit of strong coupling, as $x\rightarrow \infty$, the exact result
\begin{equation}
\label{77.2}
 E(\infty) = \frac{\pi^2}{3} \approx 3.289868,
\end{equation}
is known, called the Tonks-Girardeau limit. 

The results of calculations by different methods are shown in Table \ref{Table 13}. 

To illustrate the case of a one-dimensional Bose gas we present the results obtained 
from the fractional integration. In the case of one-dimensional Bose gas there are two 
solutions to the optimization problem.  Equation \eqref{5} possesses a single  solution,
$B_{1}=3.08574,$ corresponding to $u_{1}=-4.7686$. Equation \eqref{6} also possesses 
a single solution, $B_{2}=4.79312,$ corresponding to $u_{2}=-4.88523$.

\begin{table}
\caption{ One-dimensional Bose gas: Amplitude  ($k=8$)
}
\label{Table 13}       
\centering
\begin{tabular}{|c|c|c|c|c|c|c|c|c|c}
\hline
$B$ & genlass$_{1}$ & genlass$_{2}$ & lasso$_{1}$ & lasso$_2$ &functional \\\hline
Mittag-Leffler&3.51951 &3.51951&3.51951 &3.51951 & --\\\hline
Borel-Leroy &--&--  &-- &--&--\\ \hline
Fract. Riemann &2.59989&2.59989&2.59989&2.59989&4.50635\\\hline
Fract. Integral&\bf{3.08574}&4.79312 &\bf{3.08574}&\bf{3.08574}&4.50604\\\hline
\end{tabular}
\end{table}

Simple Borel summation is not very good in this case, while original iterated roots 
perform well, giving $B=3.52695$. Such a good performance of the iterated roots leads 
to a good result of Mittag-Leffler summation. By adding two more trial coefficients 
the results can be improved even further \cite{fr}. For instance, the method of 
fractional integral summation improves in $10$th order, giving $B=3.14517$. The latter 
results are significantly better than $B=3.81457$, produced by the Fractional Borel 
summation \cite{fr}. Borel-Leroy summation in all cases including the functional method 
gives negative results. Mittag-Leffler summation combined with the functional method 
also produces negative results.

\subsection{Membrane: Pressure.}
\label{MEM}

In the case of a two-dimensional membrane, its pressure can be calculated by
perturbation theory with respect to the wall stiffness characterized by the dimensionless 
parameter $g$ \cite{Kas2006}, so that
\be
\label{8.8}
p_k(g) = \frac{\pi^2}{8g^2} \left ( 1 + \sum_{n=1}^k a_n g^n \right ) \;  ,
\ee
with the coefficients
$$
a_1 = \frac{1}{4} \; , \qquad a_2 = \frac{1}{32} \; , \qquad
a_3=2.176347\times 10^{-3} \; ,
$$
$$ a_4=0.552721\times 10^{-4} \; , \qquad 
a_5=-0.721482\times 10^{-5} \; , \qquad 
a_6=-1.777848\times 10^{-6} \; .
$$
The so-called rigid-wall limit corresponds to infinite $g$. Employing Monte Carlo 
simulations \cite{Gom1989} gives
\be
\label{8.9}
 p(\infty) =0.0798 \pm 0.0003 \;  .
\ee

The results of calculations by different methods are shown in Table \ref{Table 14}. 

To illustrate the case of a membrane pressure  we present the results obtained from 
the fractional integration. In the case of a membrane pressure there are two very 
close solutions to the optimization problem. Equation \eqref{5} possesses a single  
solution, $B_{1}=0.0766022,$ corresponding to $u_{1}=-4.72832$. Equation \eqref{6} 
also possesses a single solution, $B_{2}=0.076647,$ corresponding to $u_{2}=-4.83955$.

\begin{table}
\caption{ Membrane Pressure: Amplitude  ($k=6$)
}
\label{Table 14}       
\centering
\begin{tabular}{|c|c|c|c|c|c|c|c|c|c}
\hline
$B$ & genlass$_{1}$ & genlass$_{2}$ & lasso$_{1}$ & lasso$_2$ &functional \\\hline
Mittag-Leffler&0.027276 &0.070684&0.044611 &0.044611 &0.059829\\\hline
Borel-Leroy &0.065546&0.065546  &0.064646  &0.065546 &0.05978\\ \hline
Fract. Riemann &\bf{0.080764}&\bf{0.080764}&\bf{0.080764}&\bf{0.080764}&0.059829\\\hline
Fract. Integral&0.076602&0.076602 &0.076647&0.076647&0.059829\\\hline
\end{tabular}
\end{table}

In this case, only fractional methods combined with generalized lasso and lasso criteria 
work well, and appear to be in good agreement with the best results from the methods 
presented in \cite{fr}. Both pure Borel summation and original iterated roots are not 
overly good in the case of membrane pressure, determining also not very good results 
from other techniques and criteria. The problem is also very difficult for all methods 
based on Pad\'{e} approximations. Two more examples of this kind will be discussed in 
the next sections.

\subsection{Gaussian polymer}

Correlation function of the Gaussian polymer \cite{grosb}, is given in the closed form 
by the Debye--Hukel function,
\begin{equation}
\Phi(x)=\frac{2}{x}-\frac{2 (1-\exp (-x))}{x^2}.
\end{equation}
For small $x>0$, we have
\begin{equation}
\Phi(x) = 1 - \frac{x}{3} +\frac{x^2}{12} -\frac{x^3}{60} + \frac{x^4}{360} +O(x^5) \; ,
\end{equation}
and 
$$
\Phi(x)= 2 x^{-1}+O(x^{-2}) \; ,
$$ 
as $x \to +\infty$. The example is quite difficult for the Pad\'{e} approximants. Their 
application leads to the so-called ``divergence by oscillation'' \cite{cor}.

The results of calculations by different methods are shown in Table \ref{Table 15}. 

To illustrate the case of a Gaussian polymer we present the results obtained from the 
fractional integration. In the case of Gaussian polymer there are altogether six 
solutions to the optimization problems. Equation \eqref{5} possesses three solutions,
$$B_{1}=1.93634,\,\,\,\,B_{2}=2.17141,\,\,\,\,B_{3}=1.80271,$$ corresponding to 
$$u_{1}=-2.831171,\,\,\,\,u_{2}=-4.453994,\,\,\,\,u_{3}=-5.58516.$$

Equation \eqref{6} also possesses three  solutions,
$$B_{4}=1.93178,\,\,\,\,B_{5}=2.18086,\,\,\,\,B_{6}=1.8024,$$ corresponding to 
$$u_{4}=-2.527555,\,\,\,\,u_{5}=-4.35428,\,\,\,\,u_{6}=-5.564997.$$

\begin{table}
\caption{Gaussian Polymer: Amplitude  ($k=11$)
}
\label{Table 15}       
\centering
\begin{tabular}{|c|c|c|c|c|c|c|c|c|c}
\hline
$B$ & genlass$_{1}$ & genlass$_{2}$ & lasso$_{1}$ & lasso$_2$ &functional \\\hline
Mittag-Leffler&1.96426 &\bf{1.97148}&1.96426 &1.96426 &1.96718\\\hline
Borel-Leroy &1.95046&1.95046  &1.95046  &1.95046 &2.48132\\ \hline
Fract. Riemann &1.85218&1.85218&1.85218&1.85218&2.07931\\\hline
Fract. Integral&1.93178&1.93178 &1.80271&1.80271&2.17141\\\hline
\end{tabular}
\end{table}

The best results are achieved by the Mittag-Leffler summation, being practically the same 
as $B=1.96978$, which follows in the $11$th order from the Fractional Borel summation 
\cite{fr}. Pure Borel summation also gives a reasonable value $B= 2.079$.

\subsection{Wilson Loop}

The $N = 4$ Super Yang--Mills Circular Wilson Loop \cite{Banks} is given by the following 
expression
\begin{equation}
\label{loop}
\Phi(y)=\frac{2 \exp \left(-\sqrt{y}\right) I_1\left(\sqrt{y}\right)}{\sqrt{y}} \; ,
\end{equation}
where $I_{1}$ is a modified Bessel function of the first kind. Let us set $\sqrt{y}=x$. 
For small $x>0$,
\begin{equation}
 \Phi(x)= 1-x+\frac{5 x^2}{8}-\frac{7 x^3}{24}+\frac{7 x^4}{64}+O(x^5) \; ,
 \end{equation} 
 and 
 $$
\Phi(x)\simeq \sqrt{\frac{2}{\pi }} x^{-\frac 32}\approx 0.797885 x^{-\frac 32} \; ,
$$ 
as $x\rightarrow \infty$. The example is very difficult for the Pad\'{e} approximants. 
Their application leads to the divergent sequences for the amplitudes \cite{cor}.

The results of calculations by different methods are shown in Table \ref{Table 16}. 
To illustrate the case of the Wilson loop, we present the results obtained from the 
fractional integration. In the case of the Wilson loop, there are altogether six 
solutions to the optimization problems. Equation \eqref{5} possesses three  solutions,
$$B_{1}=0.788166,\,\,\,\,B_{2}=0.871278,\,\,\,\,B_{3}=0.705726,$$ corresponding to 
$$u_{1}=-4.383716,\,\,\,\,u_{2}=-5.802686,\,\,\,\,u_{3}=-6.9968.$$

Equation \eqref{6} also possesses three solutions,
$$B_{4}=0.783575,\,\,\,\,B_{5}=0.871469,\,\,\,\,B_{6}=0.705646,$$ corresponding to 
$$u_{4}=-3.930875,\,\,\,\,u_{5}=-5.774946,\,\,\,\,u_{6}=-7.011296.$$

\begin{table}
\caption{Wilson Loop: Amplitude  ($k=11$)
}
\label{Table 16}       
\centering
\begin{tabular}{|c|c|c|c|c|c|c|c|c|c}
\hline
$B$ & genlass$_{1}$ & genlass$_{2}$ & lasso$_{1}$ & lasso$_2$ &functional \\\hline
Mittag-Leffler&0.813797 &0.813797&0.813797 &0.813797 &--\\\hline
Borel-Leroy &0.817901&0.817901&0.817901  &0.817901 &1.39674\\ \hline
Fract. Riemann &0.730468&0.730468&0.730468&0.730468&0.90277\\\hline
Fract. Integral&\bf{0.783575}&\bf{0.783575} &0.705646&0.705646&0.902261\\\hline
\end{tabular}
\end{table}

According to the functional method, the best results achieved are practically the same 
as $B=0.90226$ from the pure Borel summation. Fractional integration allows to get a 
little better results than $B\approx 0.8083$, which follows from the Fractional Borel 
summation \cite{fr}.

\subsection{Hard-Disc Fluid}

The fluid of hard discs of diameter $a_s$, which approximately equals the scattering
length for these objects, is an important model often used as a realistic approximation
for systems with more complicated interaction potentials. The equation of state connects
pressure $P$, temperature $T$, and density $\rho$. One often considers the ratio
\be
\label{F_1}
 Z = \frac{P}{\rho T} \; ,
\ee
where the Planck constant is set to one, which is called compressibility factor. This
factor is studied as a function of the packing fraction, or filling,
\be
\label{F_2}
f \equiv \frac{\pi}{4} \;\rho\; a_s^2  \;  .
\ee
It is known \cite{Mulero_21,Santos_22} that the compressibility factor exhibits
critical behavior at the filling $f_c = 1$, where
\be
\label{F_3}
 Z ~ \propto ~ ( f_c - f)^{-\bt} \qquad ( f \ra f_c -0 ) \; .
\ee
The exponent $\bt$ has not been calculated exactly, but it is conjectured
\cite{Mulero_21,Santos_22} to be around $\beta = 2$.

The compressibility factor for low-density has been found \cite{Clisby_23,Maestre_24}
by perturbation theory as an expansion in powers of the filling $f$. Nine terms of
this expansion are available:
$$
Z \simeq 1 + 2f + 3.12802 f^3 + 4.25785 f^3 + 5.3369 f^4 + 6.36296 f^5 +
$$
\be
\label{F_4}
 + 7.35186 f^6 +
8.3191 f^7 + 9.27215 f^8 + 10.2163 f^9 \;  ,
\ee
where $f \ra 0$. In order to reduce the consideration to the same type of problems
as treated above, we make the substitution
\be
\label{F_5}
f = \frac{x}{1+x} \; f_c 
\; ,
\quad
x = \frac{f}{f_c-f} 
.
\ee
Then, the compressibility factor at the critical point behaves as
\be
\label{F_7}
Z ~ \propto ~ x^\bt \qquad ( x \ra \infty) \; .
\ee
While the compressibility factor $Z$, as a function of $x$
becomes
$$
Z \simeq 1 + 2x + 1.12802 x^2 + 0.00181 x^3 - 0.05259 x^4 +
$$
\be
\label{F_8}
  + 0.05038 x^5 - 0.03234 x^6
+ 0.01397 x^7 - 0.0033 x^8 + 0.00618 x^9 \;   ,
\ee
where $x \ra 0$. 
Thus the problem now consists in the prediction of the large-variable
exponent $\beta$, being based on the small-variable expansion (\ref{F_8}).

Let us calculate the critical index $\bt$ for the compressibility factor of the 
hard-disc fluid, as formulated in \cite{YGB}. The results of calculations by different 
diff-log methods are shown in Table \ref{Table 17}. More  details on application 
of diff-log techniques can be found in \cite{bo,YGB}.

To illustrate the case of the hard-disk fluid, we present the results obtained from 
the diff-log fractional integration. In the case of the hard-disc fluid, there are two 
very close solutions to the optimization problem. Equation \eqref{5} possesses a single  
solution, $$\beta_{1}=1.80465,$$ corresponding to $u_{1}=-1.17396$.
Equation \eqref{6} also possesses a single  solution,
$$\beta_{2}=1.79963,$$ corresponding to $u_{2}=-0.721369$.

\begin{table}
\caption{Hard-Disc Fluid: Index  ($k=9$)
}
\label{Table 17}       
\centering
\begin{tabular}{|c|c|c|c|c|c|c|c|c|c}
\hline
$\beta$ & genlass$_{1}$ & genlass$_{2}$ & lasso$_{1}$ & lasso$_2$ &functional \\\hline
Diff-Log Mittag-Leffler&1.83516 &1.83522&1.83516 &1.83516 &1.83517\\\hline
Diff-Log Borel-Leroy &1.7852&1.7852&1.79197  &1.79197 &2.16601\\ \hline
Diff-Log Fract. Riemann &1.79208&1.79208&1.79208&1.79208&2.24014\\\hline
Diff-Log Fract. Integral&1.79963&1.79963 &1.80465&1.80465&1.80968\\\hline
\end{tabular}
\end{table}

Diff-log transformation with iterated roots gives $\bt=1.87791$, when all terms from 
the expansion are employed. Similar approach with even Pad\'{e} approximants introduced 
in \cite{pa}, gives  $\bt=1.86468$.

Method of modified diff-log Pad\'{e}-Borel summation \cite{pa}, gives comparable results, 
$\beta=2.1123$, with all terms from the original expansion being employed. Slightly 
lower results, $\beta=1.96243$, are obtained with  seven terms from the expansion. 
Various methods based on factor approximations predict critical index  
$\beta=1.884\pm 0.02$ \cite{YGB}. Thus, the functional method is in better agreement with
other estimates for the index $\bt$ than all other results presented above in the 
Table \ref{Table 17}.

\section{Comments}

We have advanced several new approaches to deal with the non-uniqueness of solutions for 
the control parameters in the optimization procedure, such as the minimal derivative and 
minimal difference conditions. The approaches typically consider the solutions of both 
optimization problems, after which all solutions for control functions are analyzed in 
the search of a minimal cost functional.

Only one-parameter Borel-type procedures are considered, since multi-parameter 
minimization procedures are essentially more complicated \cite{fr}. Two novel 
transformations are proposed using fractional integrals and fractional derivatives.

The most reliable strategy in dealing with the non-uniqueness of the solutions to the 
optimization problem is to apply the fractional integral summation with GenLass$_1$ 
selection criterion. It could be complemented if needed with the ridge functional with 
penalty imposed for the deviation from the Borel transform solution.

GenLass$_1$ selection criterion finds how strongly the transformed coefficients deviate 
from the  Borel-transformed coefficients. It not always selects the best possible solution, 
but in the overwhelming majority of considered examples leads to good accuracy, close 
to the best.

Lasso criteria are non-parametric, while GenLass$_1$ is semi-parametric, The latter 
statement means that the choice of the Borel-transformed coefficients is reasonable but 
can be different. For instance, one can measure how strongly the transformed coefficients 
deviate from the original coefficients by means of the measure GenLass$_2$. The two 
measures lead to close results but GenLass$_1$ seems to be a bit more robust. 

Selection based on the objective functional with penalties is parametric, depending 
on the optimization parameters continuously, and requiring to find the parameters at 
the last step of the optimization. 

To summarize, the problem of multiple solutions for control functions is studied.
The suggested methods of using cost functionals allow to resolve this annoying problem 
and provide rather close to each other results, enjoying reasonably good accuracy. 
Fortunately, in all physical cases, studied in the paper, all solutions to the 
optimization problems are enumerated with relative ease. The number of solutions 
in the considered examples does not exceed eight. Of course, a global optimization 
condition would seem to be preferable. It is exactly such a global optimization 
condition that is formulated at the end of Sec. 5, being based on the direct 
minimization of a cost functional constructed in the spirit of Tikhonov. Indeed, the 
use of this functional looks easier and the comparison with the methods employing 
optimization conditions shows that in some cases the results are more accurate. We 
plan to concentrate on the use of global functionals in our following work.

\vskip 2mm

{\bf Statements and Declarations}

\vskip 2mm

{\bf Author Contributions}: Conceptualization, S.G. and V.I.Y.; methodology, S.G. and V.I.Y.
All authors have read and agreed to the published version of the manuscript.

\vskip 2mm

{\bf Funding}: This research received no external funding.

\vskip 2mm

{\bf Data Availability Statement}: No new data were created or analyzed in this study. 
Data sharing is not applicable to this article.

\vskip 2mm

{\bf Conflicts of Interest}: The authors declare no conflict of interest.

\vskip 2mm

{\bf Ethical approval}: There are no studies involving human participants or animals done 
by the authors in this article.

\end{document}